\documentclass[
pre,
showpacs,preprintnumbers,
 amsmath,amssymb,twocolumn,floatfix
]{revtex4-1}

\usepackage{graphicx}% Include figure files
\usepackage{dcolumn}% Align table columns on decimal point
\usepackage{bm}% bold math
\usepackage{dsfont,bbold,subcaption}
\usepackage[utf8]{inputenc}

\newcommand{\tr}{\text{\,tr\,}}

\newcommand{\1}{\mathds{1}_n}

\begin{document}

\title{Random matrix ensembles involving Gaussian Wigner and Wishart matrices,\\ and biorthogonal structure}% Force line breaks with \\

\author{Santosh Kumar}
 \email{skumar.physics@gmail.com}
 \affiliation{Department of Physics, Shiv Nadar University, Gautam Buddha Nagar, Uttar Pradesh -- 201314, India}%Lines break automatically or can be forced with \\

\begin{abstract}
We consider four nontrivial ensembles involving Gaussian Wigner and Wishart matrices. These are relevant to problems ranging from multiantenna communication to random supergravity. We derive the matrix probability density, as well as the eigenvalue densities for these ensembles. In all cases the joint eigenvalue density exhibits a biorthogonal structure. A determinantal representation, based on a generalization of Andr\'{e}ief's integration formula, is used to compactly express the $r$-point correlation function of eigenvalues. This representation circumvents the complications encountered in the usual approaches, and the answer is obtained immediately by examining the joint density of eigenvalues. We validate our analytical results using Monte Carlo simulations.
\end{abstract}

\maketitle
%\tableofcontents

\section{Introduction}
\label{Sec1}

Wigner and Wishart matrices have been the cornerstones of random matrix theory. They find numerous applications in varied fields of knowledge~\cite{Mehta2004,Forrester2010,Handbook2011}. The inception of Wigner matrices was due to Wigner who investigated some special large dimensional random matrices to predict properties of the eigenfunctions and eigenvalues of complicated quantum mechanical systems, in particular heavy nuclei~\cite{Wigner1955a,Wigner1958}. It turns out that certain spectral characteristics of these matrices, such as {\it semicircular} distribution of eigenvalues, are universal and in fact shared by a wider class of matrices which are now collectively referred to as Wigner matrices. See, for example, Refs.~\cite{Erdos2011,TV2012} for recent surveys. An important family of Wigner matrices is realized when the matrix elements are taken as Gaussian random variables. The resulting ensembles are referred to as Gaussian ensembles~\cite{Mehta2004,Forrester2010,Dyson1972}. In the present work we use the terms Wigner and Gaussian Wigner interchangeably to mean complex Wigner matrices with Gaussian entries, more specifically matrices from the Gaussian unitary ensemble~\cite{Mehta2004,Forrester2010,Dyson1972}. Wishart matrices predate even Wigner matrices and have their origin in the field of multivariate statistics. These were introduced by Wishart who derived the generalized product-moment distribution in normal multivariate population samples~\cite{Wishart1928}. This distribution is now referred to as the Wishart distribution. In what follows, we will be concerned with ensembles comprising complex Wishart matrices.

While Wigner and Wishart matrices themselves offer plenty of aspects to explore, interestingly, various combinations of these also turn out to be of crucial importance. Many such matrix ensembles have their origin in the area of multivariate statistics~\cite{GN1999,Anderson2003,Muirhead2005}. A classic example is the Jacobi (MANOVA) ensemble which incorporates two Wishart matrices in a nontrivial manner, and arises in the problems of quantum conductance~\cite{SM2006,Forrester2006,SSW2008,KSS2009,KP2010a,KP2010b} and multiple channel fiber optics communication~\cite{DFS2013,KMV2014}. Remarkably, it also pops up in something as remote as a microscopic model of bus transport system~\cite{BBDS2006}. In recent years several other important matrix models have been explored. Some notable examples include product of complex Ginibre matrices~\cite{AIK2013,AKW2013,KZ2014,WZCT2014,AI2015,ZWSMHC2015,Kieburg2015}, Cauchy-Lorentz~\cite{Kieburg2015,Brouwer1995,BJJNPZ2002,MMSV2014,KKG2014,WWKK2015}, sum involving Wigner and Wishart matrices~\cite{MMW2012, PW2014,LMM2014,Kumar2014,KGGC2015,CKW2015}, and product of truncated unitary matrices~\cite{ABKN2014}. In addition to their natural connection with multivariate statistics, these are of interest to the fields of telecommunication~\cite{AIK2013,KGGC2015,ZWSMHC2015}, finance~\cite{BJNPZ2003,BJNPZ2004}, and random supergravity theory~\cite{MMW2012,PW2014,LMM2014}. 

In the present work we proceed further in exploring such exotic ensembles and consider four important matrix models involving Wigner and Wishart matrices. The first one comprises a \emph{ratio} involving two Wishart matrices, the second one consists of the weighted sum of a Wigner matrix and a Wishart matrix, the third is the product of a Wigner matrix and a Wishart matrix, and the fourth one embodies the weighted sum of two Wishart matrices. We derive the probability density function for these matrices, and then work out the eigenvalue statistics. The joint density of eigenvalues for these matrix models exhibit biorthogonal structure.  A determinantal representation, based on a generalization of Andr\'{e}ief's integration formula~\cite{Andreief1883,KG2010a,KG2010b}, is used to compactly express the $r$-point correlation function for all these ensembles.

The rest of the paper is organized as follows. We start with a brief discussion of biorthogonal ensembles in Sec.~\ref{Sec2}, and present the result for $r$-level correlation function for eigenvalues. Sections~\ref{Sec3}--\ref{Sec6} are devoted to the exact results for the above mentioned matrix ensembles which involve nontrivial combinations of Wigner and Wishart matrices. We conclude in Sec.~\ref{Sec7} with a brief summary and outlook. Some relevant derivations are presented in the Appendices.

% SECTION 2
\section{Biorthogonal ensembles}
\label{Sec2}

Biorthogonal ensembles arise naturally in the study of eigenvalue statistics of two matrix models~\cite{BE2006,Bertola2007}. Moreover, matrix ensembles with a unitary invariance breaking external source also give rise to biorthogonal structure~\cite{BK2004,BH1996,BH1998}. These ensembles exhibit rich mathematical structure and, at the same time, find applications in several important problems which range from quantum transport to multiple antenna telecommunication, to two-dimensional gravity~\cite{Muttalin1995,Frahm1995,Borodin1998,DF2008,ZCW2009,AIK2013,AKW2013,KZ2014,WZCT2014,Zhang2015, DKK1993}. 

We are interested here in biorthogonal ensembles of the Borodin type~\cite{Borodin1998}, which possess the following structure for joint density of its eigenvalues $\{\lambda\}$ ($\equiv\{\lambda_1,...,\lambda_n\}$):
\begin{equation}
\label{biortho}
P(\{\lambda\})=C \Delta_n(\{\lambda\})\prod_{l=1}^n w(\lambda_l) \cdot \left|f_j(\lambda_k)\right|_{j,k=1,...,n} .
\end{equation}
Here $w(\lambda)$ is a \emph{well-behaved} weight function in the desired domain, and $|\,\_\,|$ represents determinant. Also, $\Delta_n(\{\lambda\})$ is the Vandermonde determinant, 
\begin{equation}
\Delta_n(\{\lambda\})=|\lambda_k^{j-1}|_{j,k=1,...,n} =\prod_{j>k}(\lambda_j-\lambda_k).
\end{equation}
 The normalization factor $C$ follows by expanding the determinants and performing the integrals. The ensuing expression can again be represented as a determinant, as asserted by Andr\'{e}ief identity~\cite{Andreief1883}. We have
 \begin{equation}
 C^{-1}=n!\,|h_{j,k}|_{j,k=1,...,n} ,
 \end{equation}
 where
 \begin{equation}
\label{hjk}
h_{j,k}=\int d\lambda\, w(\lambda)f_j(\lambda)\,\lambda^{k-1}.
\end{equation}
For the special case of $f_j(\lambda_k)=\lambda_k^{j-1}$, we obtain the joint probability density of eigenvalues for a unitary random matrix ensemble. We note that if we replace $\Delta_n(\{\lambda\})$ by some other determinant $\left|g_j(\lambda_k)\right|$, then we have the most general form of biorthogonal ensemble, as defined by Borodin~\cite{Borodin1998}. The approach for calculating correlation function, as discussed below, extends to these as well. 

We would like to remark that the biorthogonal ensemble of Borodin type emerges after integrating out one set of eigenvalues (corresponding to one of the matrices) from the joint probability density of eigenvalues for two-matrix model; see for example Appendix~\ref{AppB}.

The $r$-point correlation function $(1\leq r\leq n)$ corresponding to Eq.~\eqref{biortho} is defined as~\cite{Mehta2004}
\begin{equation}
\label{corrdef}
R_r(\lambda_1,...,\lambda_r)=\frac{n!}{(n-r)!}\int d\lambda_{r+1}\cdots \int d\lambda_{n}\, P(\{\lambda\}).
\end{equation}
The evaluation of this correlation function usually relies on the explicit construction of biorthogonal polynomials. In~\cite{Borodin1998} a recipe has been provided to write down the correlation function in terms of a determinant of a $r$-dimensional matrix with entries containing certain two-point kernel. However, it requires inversion of a matrix. 

In the following we use a generalization of Andr\'{e}ief's integration formula to express the $r$-point correlation function in terms of the determinant of an $(n+r)$-dimensional matrix~\cite{KG2010a,KG2010b}:
\begin{align}
\label{corrfunc}
\nonumber
R_r(\lambda_1,...,\lambda_r)=(-1)^r n! \,C \prod_{l=1}^r w(\lambda_l) \\
\times \begin{vmatrix}   \mathbb{0} &  [\lambda_j^{k-1}]_{\substack{j=1,...,r\\k=1,...,n}}  \\  [f_j(\lambda_k)]_{\substack{j=1,...,n\\k=1,...,r}}  & [h_{j,k}]_{\substack{j=1,...,n\\k=1,...,n}} \end{vmatrix}.
\end{align}
In the above expression $\mathbb{0}$ represents $r\times r$ block with all entries 0, and $f_j(\lambda_k),h_{j,k}$ are as appearing in Eqs.~\eqref{biortho} and~\eqref{hjk}, respectively. In Appendix~\ref{AppA} we provide a proof of Eq.~\eqref{corrfunc} based on mathematical induction. The above representation for correlation function altogether circumvents the complications encountered in the approaches described above, and an explicit answer is obtained at once. For small $n,r$ Eq.~\eqref{corrfunc} is advantageous in the sense that it can be readily implemented and evaluated in computational packages such as Mathematica~\cite{Mathematica}. In particular the first-order marginal density of eigenvalues, which is related to the one-point correlation function as $p(\lambda)=R_1(\lambda)/n$, is given by
\begin{equation}
\label{marginal}
p(\lambda)=-(n-1)!\, C w(\lambda) \begin{vmatrix} 0 &  [\lambda^{k-1}]_{k=1,...,n}  \\  [f_j(\lambda)]_{j=1,...,n} & [h_{j,k}]_{\substack{j=1,...,n\\k=1,...,n}} \end{vmatrix}.
\end{equation}
A similar form has been used in~\cite{RKGZ2012,Kumar2014} to express the marginal density of eigenvalues.
On the other extreme, if we consider $r=n$, then the determinant in Eq.~\eqref{corrfunc} collapses to the product of determinants $|\lambda_j^{k-1}|$ and $|f_j(\lambda_k)|$, along with the factor $(-1)^n$, and thereby yields $n! \,P(\{\lambda\})$, as expected.

As discussed in the Introduction, in the following sections we consider four matrix ensembles where such biorthogonal structure emerges. The joint density of eigenvalues for these ensembles appear in the form of Eq.~\eqref{biortho}, and hence the $r$-point correlation function can be written down immediately with the aid of Eq.~\eqref{corrfunc}.

% SECTION 3
\section{Ratio involving two Wisharts}
\label{Sec3}

%SUBSECTION 3a
\subsection{Matrix model and probability density}
We consider an ensemble of $n\times n$ dimensional complex matrices
\begin{equation}
\label{ratio}
H=(aA)(\1+bB)^{-1},
\end{equation}
where $a$ and $b$ are some non-negative scalars (for definiteness), and $A$ and $B$ are positive-definite Hermitian matrices, respectively, from the complex Wishart distributions
\begin{equation}
\label{Wisharts}
\mathcal{P}_A(A)\propto e^{-\tr A}|A|^{n_A-n},~~
\mathcal{P}_B(B)\propto e^{-\tr B}|B|^{n_B-n}.
\end{equation}
Here $n_A,n_B\ge n$ are the respective degrees of freedom for the two distributions. We may refer to the ensemble described by Eq.~\eqref{ratio} as a quotient ensemble. For $b\rightarrow0$ we have the usual complex Wishart, while the limit $a=b\rightarrow\infty$ leads to the ensemble~$AB^{-1}$, which is a multivariate generalization of the beta distribution of the second kind~\cite{Kumar2015}. We also note that $(\1+bB)^{-1/2}(aA)(\1+bB)^{-1/2}$, $aA (\1+bB)^{-1}$ and $(\1+bB)^{-1}(aA)$ share the identical nonnegative eigenvalues as they correspond to the same generalized eigenvalue problem and lead to the secular equation $|aA-\lambda (\1+bB)|=0$.
We will see below that the above construction leads to a very interesting matrix model whose probability density involves confluent hypergeometric function of the second kind (Tricomi's function) with matrix argument~\cite{JJ1985}. Moreover, this matrix model is of direct relevance to the problem of multiple antenna relay systems~\cite{TV2005}.

% FIGURE 1
\begin{figure*}[!ht]
\centering
\begin{subfigure}{.42\textwidth}
  \centering
  \includegraphics[width=\linewidth]{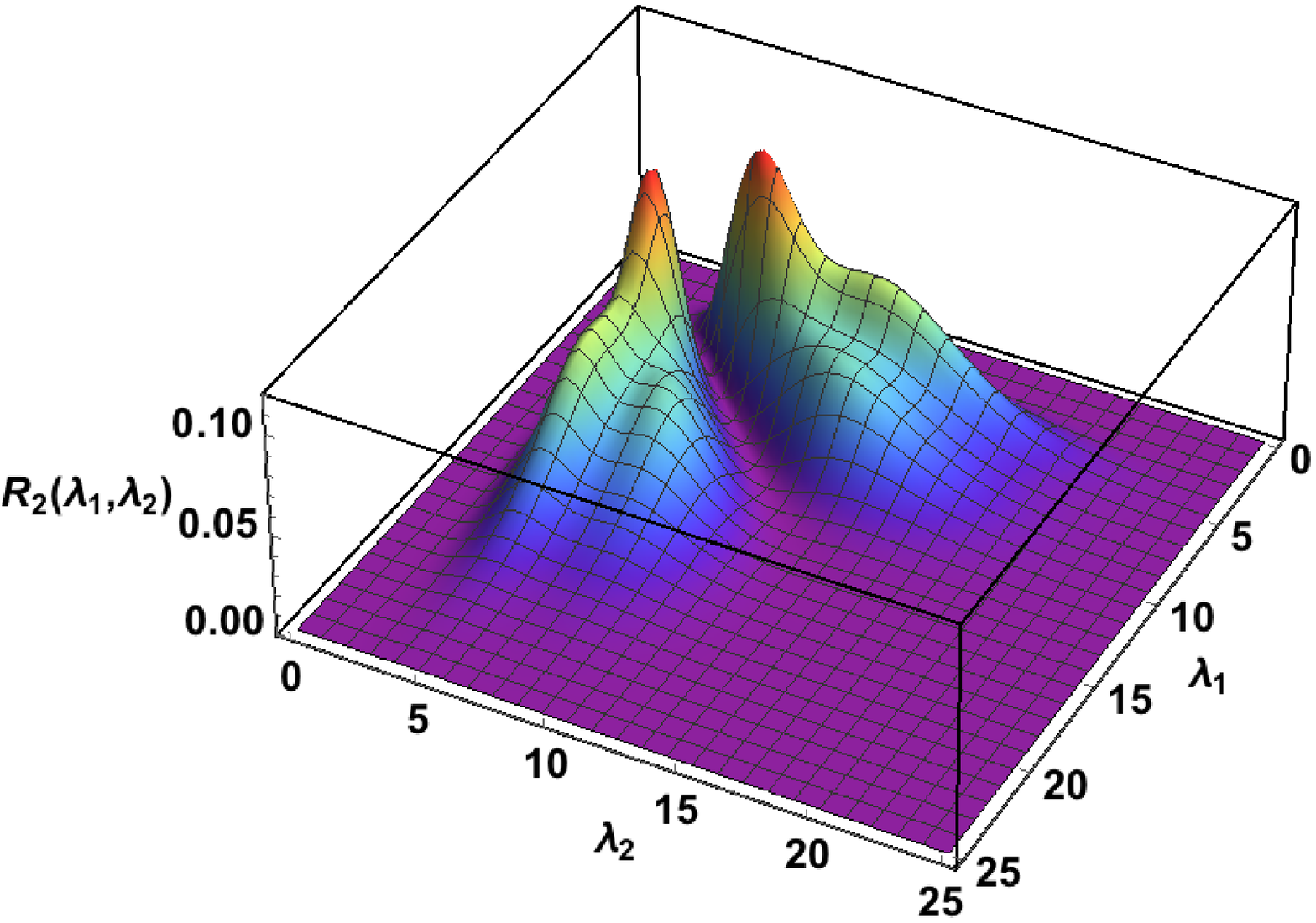}
 \caption{}
  \label{Fig1a}
\end{subfigure}
~~~~~~~
\begin{subfigure}{.42\textwidth}
  \centering
  \includegraphics[width=\linewidth]{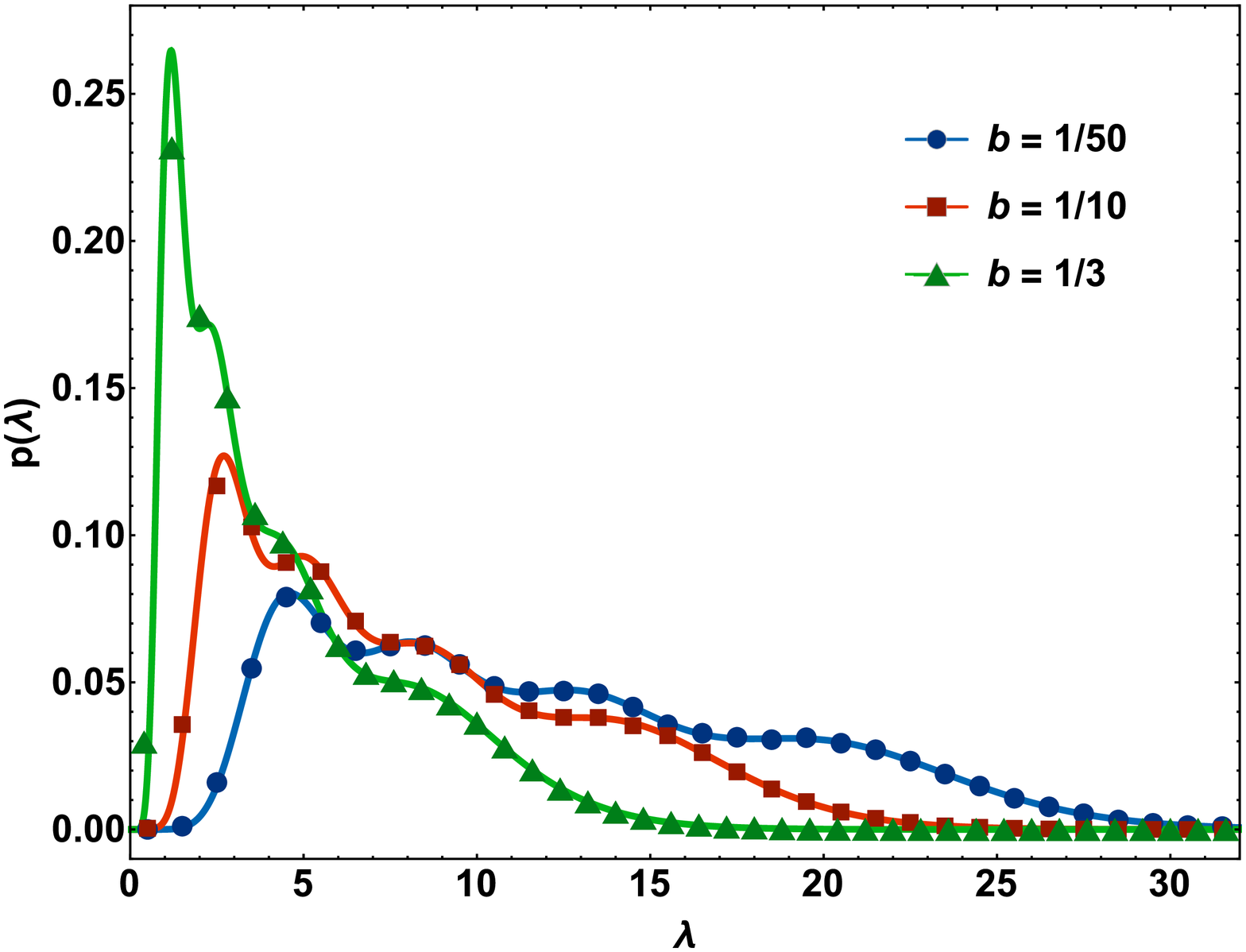}
  \caption{}
  \label{Fig1b}
\end{subfigure}
\caption{Eigenvalue densities for the ratio of Wishart matrices, Eq.~\eqref{ratio}. (a) Two point correlation function for $n=3, n_A=20,n_B=21, a=2,b=1/5$; (b) marginal density for $n=4, n_A=14, n_B=9, a=1$ and different $b$ values, as indicated. The symbols (circles, squares, triangles) shown in (b) are using numerical simulation.}
\label{Fig1}
\end{figure*}
% FIGURE 1

% FIGURE 2
\begin{figure}[ht]
  \includegraphics[width=0.85\linewidth]{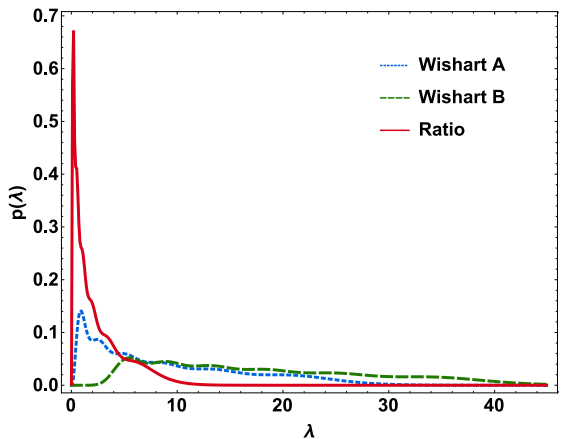}
  \caption{Marginal densities for two Wishart matrices (Wishart $A$, Wishart $B$) and the ratio, as defined in Eq.~\eqref{ratio}. The parameters are $n=6$, $n_A=9$, $n_B=18$, and $a=2, b=1/2.$\hspace{5.2cm} }
\label{Fig2}
\end{figure}
% FIGURE 2

The probability density of $H$ can be calculated as
\begin{align}
\nonumber
\mathcal{P}_H(H)=\int d[A]\mathcal{P}_A(A)\int d[B]\mathcal{P}_B(B)\\
\times \delta\big(H-(aA)(\1+bB)^{-1}\big).
\end{align}
Here the delta function with matrix argument represents the product of delta functions with scalar arguments, one for each independent real and imaginary component of $H-(aA)(\1+bB)^{-1}$. Also, $d[A]$, etc. represent the flat measure involving the product of the differentials of all independent variables occurring within the matrix.
Implementation of the Fourier representation for delta function and the cyclic invariance property of trace gives
\begin{align}
\label{FDelta}
\nonumber
\mathcal{P}_H(H)\propto \int d[K]\int d[A]\mathcal{P}_A(A)\int d[B]\mathcal{P}_B(B)\\
\times e^{i\tr KH}e^{-i\tr(aA(\1+bB)^{-1}K)}.
\end{align}
The matrix $K$ in the above equation possesses symmetry properties identical to those of $H-(aA)(\1+bB)^{-1}$.
Using Eq.~\eqref{Wisharts}, reordering the integrals, and considering the transformation $K\rightarrow (\1+bB)K$, we obtain
\begin{align}
\nonumber
&\mathcal{P}_H(H)\propto \int d[B]e^{-\tr B}|B|^{n_B-n}|\1+bB|^n\\
&\times\int d[K]e^{i\tr KH(\1+bB)}\int d[A]e^{-\tr A(\1+ia K)}|A|^{n_A-n}.
\end{align}
Integration over $A$ yields
\begin{align}
\nonumber
\mathcal{P}_H(H)&\propto \int d[B]e^{-\tr B}|B|^{n_B-n}|\1+bB|^n \\
&\times \int d[K]e^{i\tr KH(\1+bB)}|\1+ia K|^{-n_A}.
\end{align}
The $K$ integral can be identified as a variant of Ingham-Siegel-Fyodorov integral~\cite{F2002} and leads to
\begin{align}
\nonumber
\mathcal{P}_H(H)\propto \int d[B]e^{-\tr B}|B|^{n_B-n}|\1+bB|^n~~~~ \\
\times e^{-\tr a^{-1}H(\1+bB)}|H|^{n_A-n}|\1+bB|^{n_A-n}.
\end{align}
We may write
\begin{equation}
\mathcal{P}_H(H)\propto e^{-a^{-1}\tr H}|H|^{n_A-n}\,\Phi(H),
\end{equation}
where
\begin{equation}
\Phi(H)=\int d[B]e^{-\tr(\1+a^{-1}bH)B}|B|^{n_B-n}|\1+bB|^{n_A}.
\end{equation}
$\Phi(H)$ can be expressed in terms of the confluent hypergeometric function of the second kind (Tricomi's function) with matrix argument~\cite{JJ1985},
\begin{equation}
\label{PsiX}
\Psi(\alpha,\gamma;X)=\frac{1}{\Gamma_n(\alpha)}\int d[Y] e^{-\tr XY}|Y|^{\alpha-n}|\1+Y|^{\gamma-\alpha-n},
\end{equation}
as
\begin{equation}
\Phi(H)=\frac{\Gamma_n(n_B)}{b^{n n_B}}\Psi(n_B,n_A+n_B+n;(b^{-1}\1+a^{-1}H)).
\end{equation}
Here $\Gamma_n(n_B)$ is the multivariate Gamma function:
\begin{equation}
\label{multigamma}
\Gamma_n(\alpha)=\pi^{n(n-1)/2}\prod_{j=1}^n\Gamma(\alpha-j+1).
\end{equation}
Thus, we finally have the result
\begin{align}
\label{PH1}
\nonumber
\mathcal{P}_H(H)&\propto\, e^{-a^{-1}\tr H}|H|^{n_A-n}\\
&\times\Psi(n_B,n_A+n_B+n;(b^{-1}\1+a^{-1}H)).
\end{align}

%SUBSECTION 3b
\subsection{Eigenvalue statistics}
We now derive the joint density of eigenvalues for the matrix model of Eq.~\eqref{ratio}.
As implied by the result in Appendix~\ref{AppB}, $\Psi(\alpha,\gamma;X)$ of Eq.~\eqref{PsiX} admits the following determinantal representation in terms of elements involving hypergeometric function of the second kind (Tricomi's function) with scalar argument~\cite{BE1953,AS1972}:
\begin{equation}
\Psi(\alpha,\gamma;X)\propto\frac{1}{\Delta(\{x\})}|U(\alpha -j+1,\gamma-j-n+2;x_k)|_{j,k=1,...,n}.
\end{equation}
Here $x_j$'s are the eigenvalues of $X$. The  joint density of eigenvalues ($0<\lambda_1,...,\lambda_n<\infty$) of $H$, therefore, follows immediately from Eq.~\eqref{PH1}, and possesses the biorthogonal structure as in Eq.~\eqref{biortho} with
\begin{equation}
w(\lambda)=e^{-\lambda/a}\lambda^{n_A-n},
\end{equation}
\begin{equation}
f_j(\lambda_k)=U\Big(n_B -j+1,n_A+n_B-j+2;~\frac{1}{b}+\frac{\lambda_k}{a}\Big).
\end{equation}
The $h_{j,k}$ of Eq.~\eqref{hjk} is obtained as
\begin{align}
\nonumber
h_{j,k}&=a^{n_A-n+k}\Gamma(n_A-n+k)\\
&\times U\Big(n_B-j+1,n_B+n-j-k+2;~\frac{1}{b}\Big).
\end{align}
We used here the integral result
\begin{equation}
\int_0^\infty\!\!\!dz\, z^c e^{-z}\,U(a,b;z+m)=\Gamma(c+1)\,U(a,b-c-1;m),
\end{equation}
which holds whenever the integral is convergent.
Therefore, $r$-point correlation function and the marginal density follow immediately from Eqs.~\eqref{corrfunc} and~\eqref{marginal}.

In Fig.~\ref{Fig1a} we show the two-point correlation function for parameter values indicated in the caption. Although not shown here for the sake of clarity, a two-dimensional histogram obtained using Monte-Carlo simulation agrees well with the analytical plot. In Fig.~\ref{Fig1b} marginal density of eigenvalues is shown for parameter values mentioned in the caption. In this case simulation results are also depicted with the aid of symbols, and are in excellent agreement with the analytical curves. As already indicated, the parameters $a$ and $b$ give freedom to realize a variety of densities using two Wishart matrices, the exact outcome being dependent on the dimensions of the constituent matrices. As an example, in Fig.~\ref{Fig2} we show the density corresponding to the quotient ensemble defined by Eq.~\eqref{ratio} along with the densities of the constituent Wishart matrices, which can be calculated using the result
\begin{align}
\label{pW}
\nonumber
p_{\mathrm{W}}(\lambda)&=\frac{\Gamma(n)}{\Gamma(s+n)}\lambda^{s}e^{-\lambda}\\
&\times [L_{n-1}^{(s)}(\lambda)L_{n}^{(s+1)}(\lambda)-L_{n}^{(s)}(\lambda)L_{n-1}^{(s+1)}(\lambda)].
\end{align}
Here $L_\mu^{(s)}(\lambda)$ represents associated Laguerre polynomial of degree $\mu$, and the parameter $s$ is given by $n_A-n$ or $n_B-n$, i.e., it is the difference of degree of freedom and dimension of the Wishart matrix. We should underline that the resulting spectra, tunable by $a$ and $b$, can have a crucial role in deciding the behavior of metric which follow from the eigenvalue statistics, such as channel capacity and outage probability in the case of multiple access channel (MAC) and interference channel (IC) in multiple-input multiple-output (MIMO) communication~\cite{TV2005}.

% FIGURE 3
\begin{figure*}[ht]
\centering
\begin{subfigure}{.5\textwidth}
  \centering
  \includegraphics[width=\linewidth]{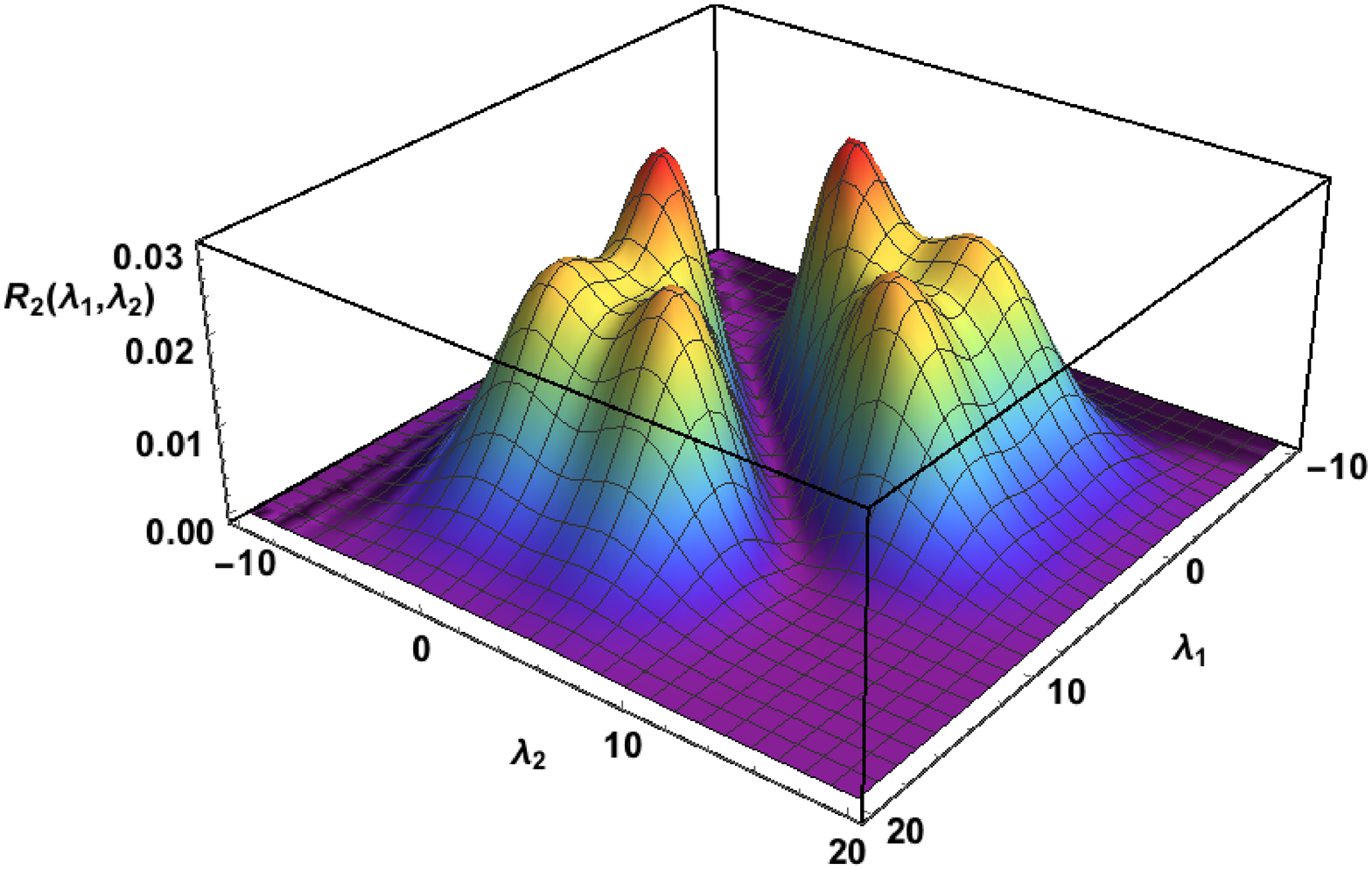}
  \caption{}
  \label{Fig3a}
\end{subfigure}
~~~~~~~
\begin{subfigure}{.41\textwidth}
  \centering
  \includegraphics[width=\linewidth]{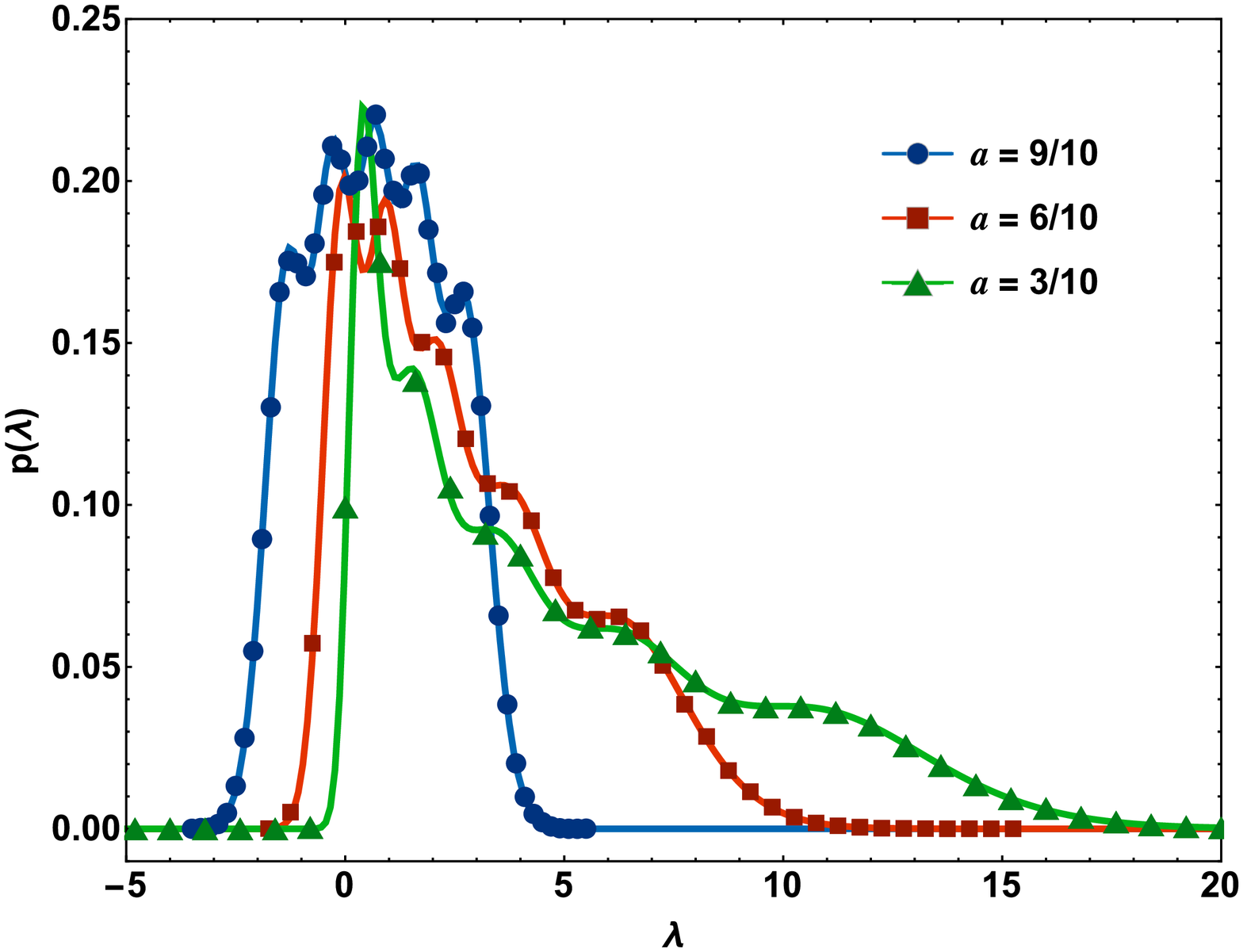}
  \caption{}
  \label{Fig3b}
\end{subfigure}
\caption{Eigenvalue densities for weighted sum of Wigner and Wishart matrices, Eq.~\eqref{WigWis}. (a) Two point correlation function for $n=3, n_B=4, a=4, b=1$; (b) marginal density for $n=5, n_B=7, b=1-a$ and $a=9/10, 6/10, 3/10$ as indicated in the figure.\hspace{14.4cm}~}
\label{Fig3}
\end{figure*}
% FIGURE 3

% FIGURE 4
\begin{figure}[ht]
  \includegraphics[width=0.85\linewidth]{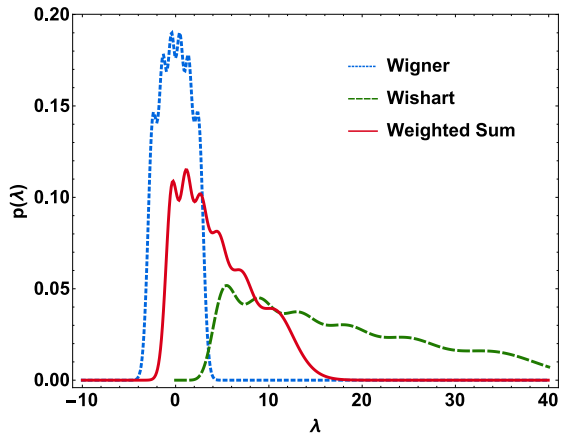}
  \caption{Marginal densities for Wigner matrix, Wishart matrix and their weighted sum, as given in Eq.~\eqref{WigWis}. The parameter values are $n=6,n_B=9,a=1,b=1/2.$ }
\label{Fig4}
\end{figure}
% FIGURE 4

% SECTION 4
\section{Weighted sum of a Wigner and a Wishart}
\label{Sec4}

%SUBSECTION 4a
\subsection{Matrix model and probability density}
We now consider an ensemble comprising weighted sum of Wigner and Wishart matrices:
\begin{equation}
\label{WigWis}
H=aA+bB.
\end{equation}
Here $A$ and $B$ are respectively $n$-dimensional Hermitian and positive-definite-Hermitian matrices from the distributions
\begin{equation}
\label{WigWisDist}
\mathcal{P}_A(A)\propto e^{-\tr A^2},~~~
\mathcal{P}_B(B)\propto e^{-\tr B}|B|^{n_B-n},
\end{equation}
and $a,b$, as before, are non-negative scalars. Also, $n_B\ge n$. For $b\rightarrow 0$, with $a>0$, we have the Wigner (Gaussian unitary) ensemble. On the other hand, for $a\rightarrow 0$, with $b>0$, we obtain the Wishart (Laguerre unitary) ensemble. Therefore, by considering $b=1-a$, and by varying $a$ between 0 and 1, we have an ensemble which interpolates between the Wishart and Wigner ensembles. To the best of our knowledge, for this matrix model only the first order marginal density of eigenvalues is known in the large $n$ asymptotic regime using the tools of free probability~\cite{Speicher1993}.  A matrix ensemble similar to that in Eq.~\eqref{WigWis} has been used to model the Hessian matrix in the context of supergravity~\cite{MMW2012,PW2014,LMM2014}. 

To obtain the probability density function for $H$ we introduce the Fourier representation of delta function as in Eq.~\eqref{FDelta}. Reordering of the integrals, and use of the cyclic invariance property of trace then gives
\begin{align}
\nonumber
\mathcal{P}_H(H)&\propto \int d[B]e^{-\tr B}|B|^{n_B-n}\int d[K]e^{i\tr(H-bB)K}\\
&\times \int d[A]e^{-\tr A^2-ia\tr KA}.
\end{align}
Evaluation of the Gaussian integral involving $A$ leads to 
\begin{align}
\nonumber
\mathcal{P}_H(H)&\propto \int d[B]e^{-\tr B}|B|^{n_B-n}\\
&\times\int d[K]e^{-\frac{a^2}{4}\tr K^2}e^{i\tr(H-bB)K}.
\end{align}
The Gaussian integral over $K$ can also be performed and yields
\begin{equation}
\label{MM2PH}
\mathcal{P}_H(H)\propto e^{-\frac{1}{a^2}\tr H^2}\Phi(H),
\end{equation}
where
\begin{equation}
\Phi(H)=\int d[B]e^{-\tr B^2}e^{\tr (\frac{2}{a}H-\frac{a}{b}\1)B}|B|^{n_B-n}.
\end{equation}

%SUBSECTION 4b
\subsection{Eigenvalue statistics}

We now calculate the eigenvalue statistics corresponding to Eq.~\eqref{MM2PH}. Using the result in Appendix~\ref{AppB} we know that $\Phi(H)$ is determined solely by the eigenvalues ($-\infty<\lambda_1,...,\lambda_n<\infty$) of $H$ as
\begin{equation}
\Phi(H)\propto \frac{1}{\Delta(\{\lambda\})}|f_j(\lambda_k)|_{j,k=1,...,n},
\end{equation}
where 
\begin{equation}
f_j(\lambda_k)=\int_0^\infty d\mu\,  \mu^{n_B-j} e^{-\mu^2+(\frac{2\lambda_k}{a}-\frac{a}{b})\mu}.
\end{equation}
This integral can be evaluated in terms of confluent hypergeometric function of the first kind (Kummer's function)~\cite{BE1953,AS1972}, and leads to the joint density, Eq.~\eqref{biortho}, with~\footnote{For $\lambda_k<a^2/2b$, a much simpler representation is possible in terms of confluent hypergeometric function of the second kind.}
\begin{widetext}
\begin{align}
\nonumber
f_j(\lambda_k)&=\frac{1}{2}\Gamma\left(\frac{n_B-j+1}{2}\right)\,_1F_1\left(\frac{n_B-j+1}{2},\frac{1}{2};\left(\frac{\lambda_k}{a}-\frac{a}{2b}\right)^2\right)\\
&+\left(\frac{\lambda_k}{a}-\frac{a}{2b}\right)\Gamma\left(\frac{n_B-j+2}{2}\right)\,_1F_1\left(\frac{n_B-j+2}{2},\frac{3}{2};\left(\frac{\lambda_k}{a}-\frac{a}{2b}\right)^2\right).
\end{align}
The weight function is read from Eq.~\eqref{MM2PH} as 
\begin{equation}
w(\lambda)=e^{-\lambda^2/a^2}.
\end{equation} 
In this case obtaining a closed form for $h_{j,k}$ requires some effort. A possible representation is in terms of hypergeometric $_2F_2$~\cite{BE1953,AS1972}:
\begin{align}
h_{j,k}=\frac{\sqrt{\pi}\,b^{n_B-j+k}}{a^{n_B-j}}\,\Gamma(n_B-j+k)\, \,_2F_2\left(\frac{1-k}{2},\frac{2-k}{2};\frac{1-n_B+j-k}{2},\frac{2-n_B+j-k}{2};\frac{a^2}{4b^2}\right).
\end{align}
With the above explicit results, the $r$-point correlation function of Eq.~\eqref{corrfunc} is readily obtained.
\end{widetext}

We show the two-point correlation function surface in Fig.~\ref{Fig3a}. The marginal density is shown along with the Monte-Carlo simulation outcome in Fig.~\ref{Fig3b}. In particular, for Fig.~\ref{Fig3b} we have considered $b=1-a$. Therefore, a crossover is seen from Wigner density (\emph{semicircle} type) to Wishart density (Mar\v{c}enko-Pastur type). In Fig.~\ref{Fig4} we compare the eigenvalue density of the composite ensemble with the eigenvalue density for the constituent Wishart ensemble given by Eq.~\eqref{pW} and that of the Gaussian Wigner ensemble evaluated using
\begin{align}
\label{pGW}
p_{\text{GW}}(\lambda)=\frac{e^{-\lambda^2}}{2^{n}\sqrt{\pi}\,n!}[H_{n}(\lambda)H_{n}(\lambda)-H_{n-1}(\lambda)H_{n+1}(\lambda)].
\end{align}
Here $H_\mu(\lambda)$ represents the Hermite polynomial of degree $\mu$. We can see that the Wishart constituent of the composite matrix tends to keep the eigenvalues in the positive half of the real line, while the Wigner part pulls them toward the negative half and tries to make the density symmetric about zero, thereby giving rise to an interesting hybrid density. 

% FIGURE 5
\begin{figure*}[t]
\centering
\begin{subfigure}{.45\textwidth}
  \centering
  \includegraphics[width=\linewidth]{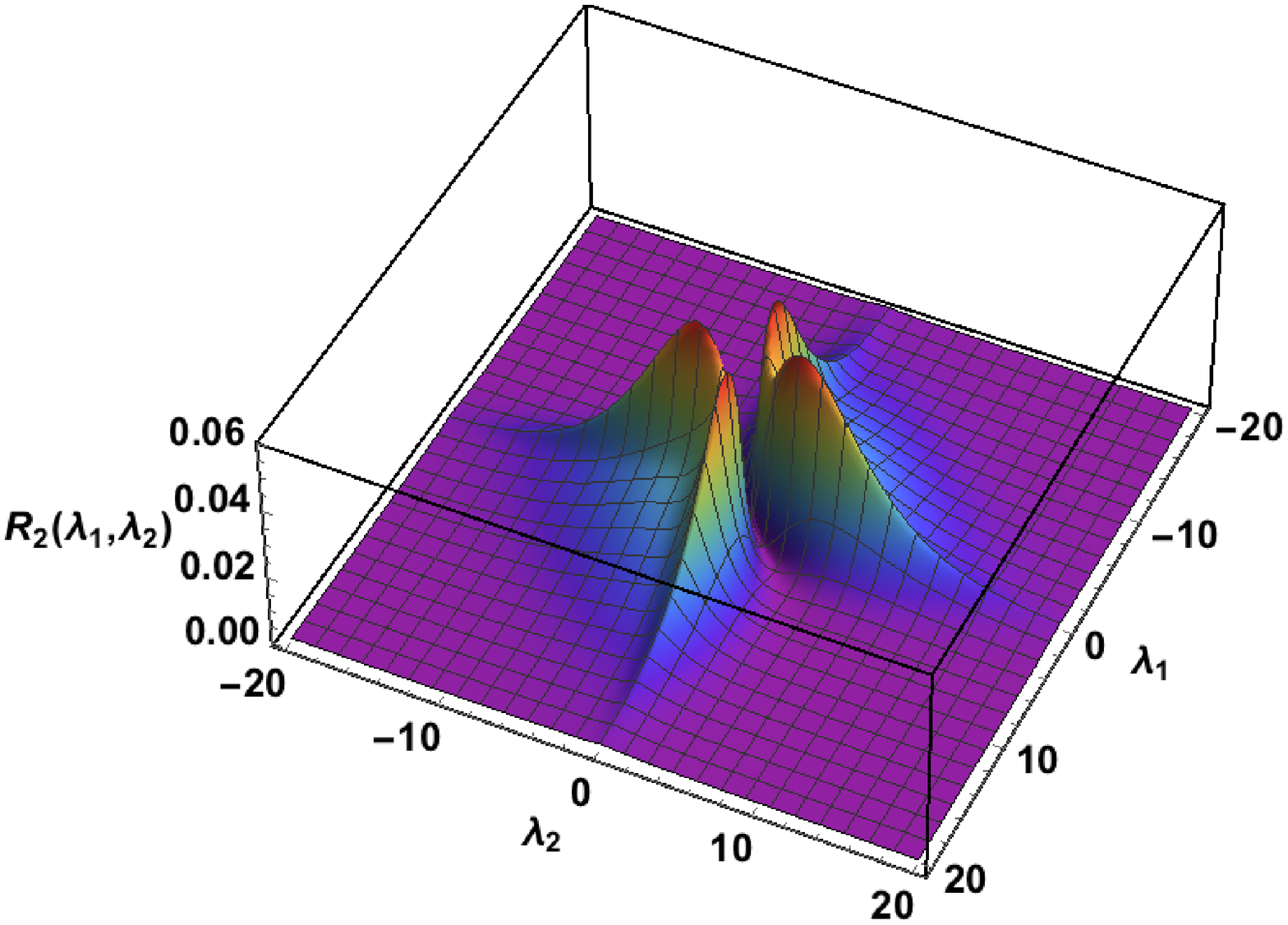}
  \caption{}
  \label{Fig5a}
\end{subfigure}
\begin{subfigure}{.41\textwidth}
~~~~
  \centering
  \includegraphics[width=\linewidth]{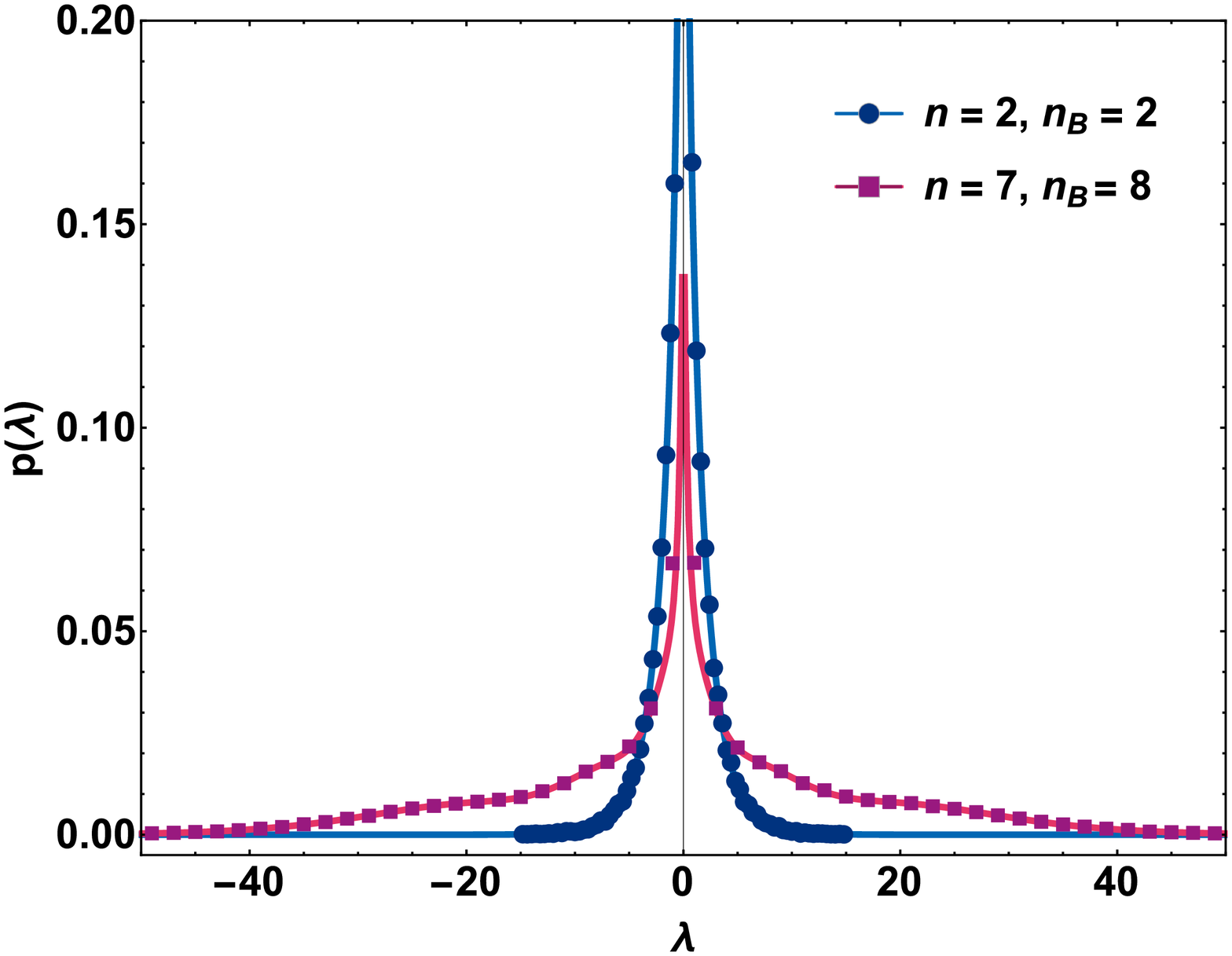}
  \caption{}
  \label{Fig5b}
\end{subfigure}
\caption{Eigenvalue densities for product of Wigner and Wishart matrices, Eq.~\eqref{WigWisProd}. (a) Two point correlation function for $n=3, n_B=5$; (b) marginal density of eigenvalues for $n=n_B=2$ and $n=7, n_B=8.$~~~~~~~~~~~~~~~~~}
\label{Fig5}
\end{figure*}
% FIGURE 5

% FIGURE 6
\begin{figure}[ht]
  \includegraphics[width=0.85\linewidth]{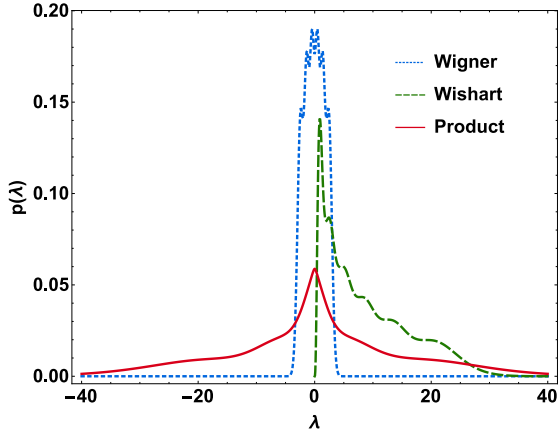}
  \caption{Marginal densities for Wigner matrix, Wishart matrix, and their product as in Eq.~\eqref{WigWisProd}. The dimen- sions of the matrices are given by $n=6, n_B=9.$~~~~~~ }
  \label{Fig6}
\end{figure}
% FIGURE 6

% SECTION 5
\section{Product of a Wigner and a Wishart}
\label{Sec5}

%SUBSECTION 5a
\subsection{Matrix model and probability density}
We now  consider an ensemble defined by
\begin{equation}
\label{WigWisProd}
H=AB,
\end{equation}
where $A$ and $B$, respectively, are Wigner and Wishart matrices from the distributions given in~\eqref{WigWisDist}. We note here that $A$ is Hermitian but $B$ is Hermitian and positive-definite as well; therefore, the signs of eigenvalues of $H$ are decided by the respective signs of eigenvalues of $A$~\cite{Serre2010}. As a consequence we expect the resultant first order marginal density to be symmetric about the origin, similar to that in the Wigner case.

We introduce the matrix delta function, as in Eq.~\eqref{FDelta}, to obtain
\begin{align}
\nonumber
 \mathcal{P}_H(H)&\propto\int d[K]\int d[A]\int d[B]e^{i\tr K(H-AB)}\\
 &~~~~~\times e^{-\tr A^2}e^{-\tr B}|B|^{n_B-n}.
\end{align}
We reorder the integrals and use the cyclic invariance of trace to get
\begin{align}
\nonumber
 \mathcal{P}_H(H)&\propto\int d[A] e^{-\tr A^2}\int d[K]e^{i\tr KH}\\
 &~~~~~\times\int d[B]e^{-\tr B(\1+i KA)}|B|^{n_B-n}.
\end{align}
Integral over $B$ can be done to give
\begin{equation}
 \mathcal{P}_H(H)\propto\int d[A] e^{-\tr A^2}\int d[K]e^{i\tr KH}|\1+i KA|^{-n_B}.
\end{equation}
We now employ the transformation $K\rightarrow KA^{-1}$, which leads to
\begin{align}
\nonumber
 \mathcal{P}_H(H)&\propto\int d[A] e^{-\tr A^2}|A|^{-n}\\
& \times \int d[K]e^{i \tr KA^{-1}H}|\1+i K|^{-n_B}.
\end{align}
The $K$-integral can now be performed~\cite{F2002} and yields
\begin{equation}
\label{PH5a}
 \mathcal{P}_H(H)\propto |H|^{n_B-n}\,\Phi(H)
\end{equation}
with 
\begin{equation}
\Phi(H)=\int d[A] e^{-\tr (A^2+A^{-1}H)}|A|^{-n_B}\,\Theta(A^{-1}H).
\end{equation}
Here $\Theta(\_)$ represents the Heaviside theta function and requires $A^{-1}H$ to be positive-definite for a non-vanishing result. 
%SUBSECTION 5b
\subsection{Eigenvalue statistics}
With a little modification the result in Appendix~\ref{AppB} implies that $\Phi(H)$ is determined by the eigenvalues ($-\infty<\lambda_1,...,\lambda_n<\infty$) of $H$ as
\begin{equation}
\Phi(H)\propto \frac{1}{\Delta(\{\lambda\})}|f_j(\lambda_k)|_{j,k=1,...,n},
\end{equation}
where 
\begin{equation}
f_j(\lambda_k)=\int_0^u d\mu\, \mu^{-n_ B+n+j-2}e^{-\mu^2-\lambda_k/\mu},
\end{equation}
with $u =-\infty$ for $\lambda < 0$ and $u = \infty$ for $\lambda > 0$.
This integral can be evaluated compactly in terms of Meijer G-function~\cite{BE1953} as
\begin{align}
\nonumber
&f_j(\lambda_k)=\frac{\lambda_k^{-n_B+n+j-1}}{2^{\,-n_B+n+j}\sqrt{\pi}}\\
&\times G^{3,0}_{0,3}\left(
\begin{array} {c}
\text{---} \\ \dfrac{n_B-n-j+1}{2},\, \dfrac{n_B-n-j+2}{2},\, 0 \end{array}\Bigg|\, \frac{\lambda_k^2}{4} \right).
\end{align}
Meijer G-functions have also appeared in the correlation kernels for product of complex Ginibre matrices~\cite{AIK2013,AKW2013,KZ2014,WZCT2014,AI2015,ZWSMHC2015,Kieburg2015}, and product of truncated unitary matrices~\cite{ABKN2014}.
The weight function $w(\lambda)$, in view of Eq.~\eqref{PH5a}, is
\begin{equation}
w(\lambda)=\lambda^{n_B-n},
\end{equation}
which leads to the following expression of $h_{j,k}$:
\begin{equation}
h_{j,k}=\frac{1+(-1)^{j+k}}{2}\,\Gamma(n_B-n+k)\,\Gamma\left(\frac{j+k-1}{2}\right).
\end{equation}
Equation~\eqref{corrfunc} now determines correlation functions of all orders for the matrix model~\eqref{WigWisProd}.

Figure~\ref{Fig5a} shows the two-point correlation function of eigenvalues, while Fig.~\ref{Fig5b} depicts the marginal density. For $n=n_B$ the density exhibits a logarithmic singularity at $\lambda=0$. This can be seen in $n=n_B=2$ plot in Fig.~\ref{Fig5b}. In Fig.~\ref{Fig6} we display the eigenvalue density for the product of Wigner and Wishart matrices, along with the densities for the constituent matrices calculated using Eqs.~\eqref{pW} and~\eqref{pGW}. Figure~\ref{Fig6} should be compared with Fig.~\ref{Fig4} where we have used matrices with dimensions same as in the present case. The distinct nature of the resultant densities in these two cases is expected because of very different characteristics of the underlying composition. The shapes of the marginal density curves here are reminiscent of the density of eigenvalues of adjacency matrices in scale free networks~\cite{JB2007,BJ2007,NR2008} and matrices defined on Poissonian random graphs~\cite{Kuhn2008}.

% FIGURE 7
\begin{figure*}[t]
\centering
\begin{subfigure}{.43\textwidth}
  \centering
  \includegraphics[width=\linewidth]{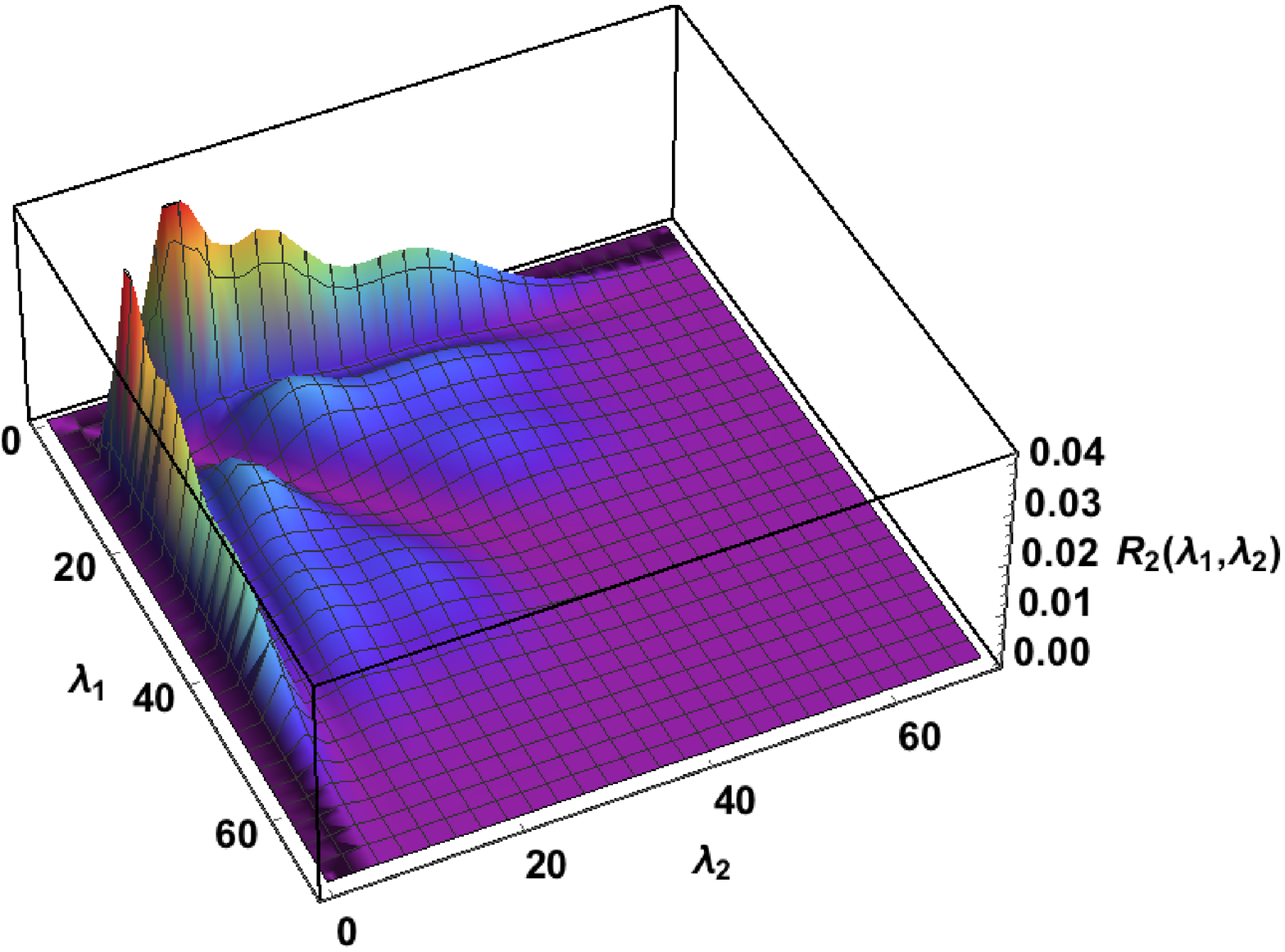}
  \caption{}
  \label{Fig7a}
\end{subfigure}
\begin{subfigure}{.41\textwidth}
  \centering
  \includegraphics[width=\linewidth]{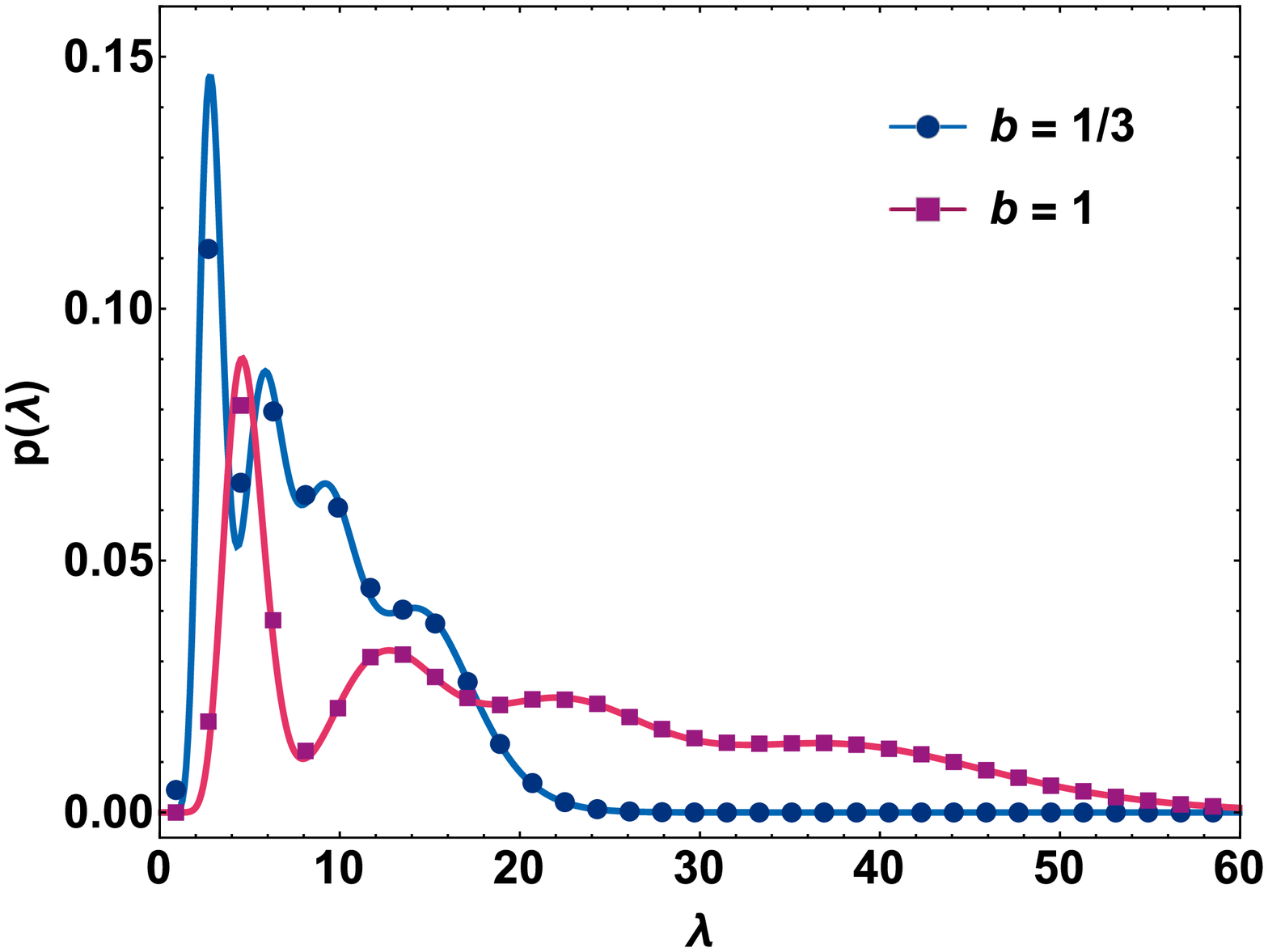}
  \caption{}
  \label{Fig7b}
\end{subfigure}
\caption{Eigenvalue densities for weighted sum of Wishart and correlated-Wishart matrices, Eq.~\eqref{SOW2}. (a) Two point correlation function for $b=1$; (b) marginal density for $b=1/3$ and 1. Common parameters for both the figures are $a=1/4, n=4, n_A=10,$ $n_B=11$, and $(\sigma_1,\sigma_2,\sigma_3,\sigma_4)=(5/2,1/3,2,7/4).$\hspace{6cm}~~~~~}
\label{Fig7}
\end{figure*}
% FIGURE 7

% FIGURE 8
\begin{figure}[ht]
  \includegraphics[width=0.85\linewidth]{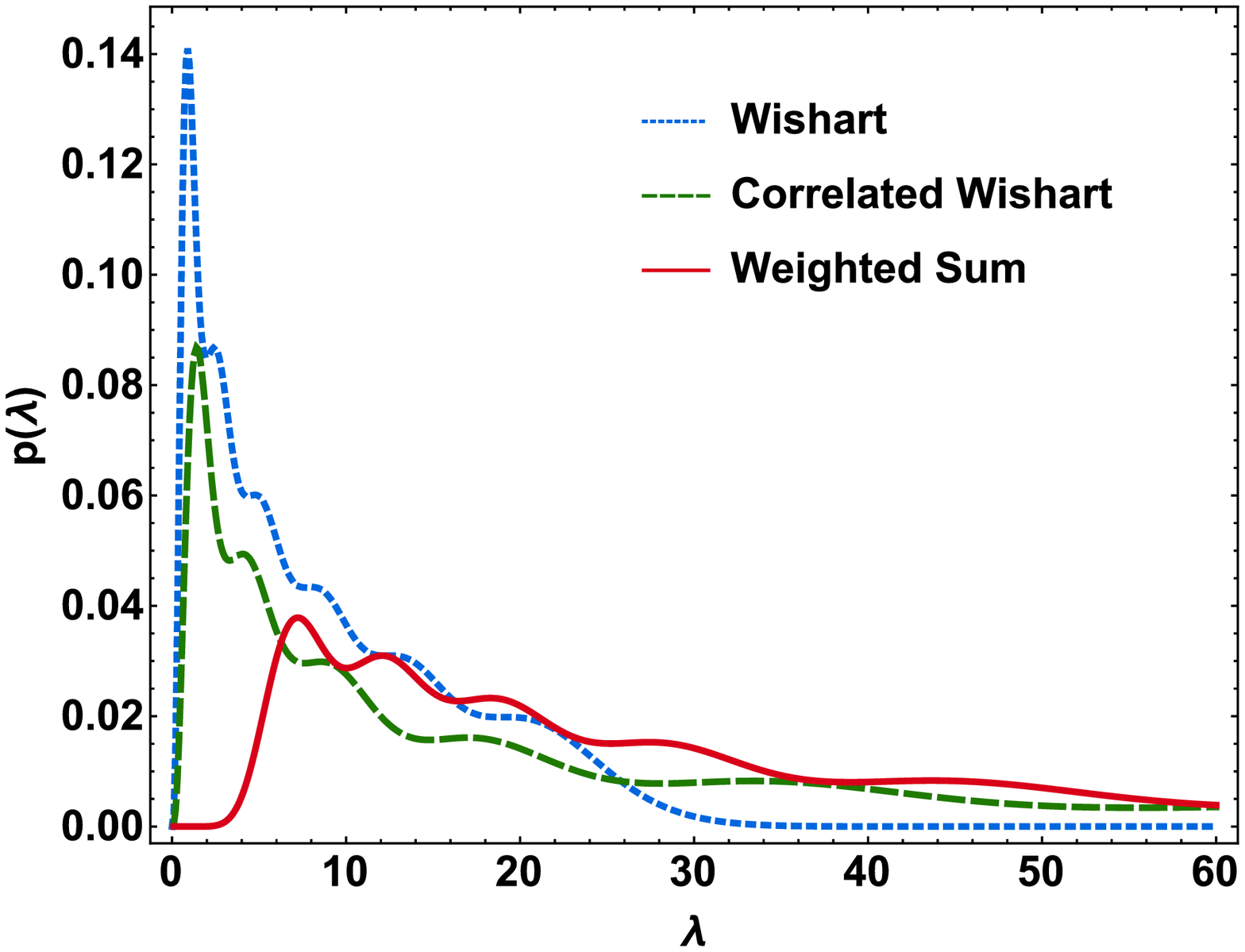}
  \caption{Marginal densities for Wishart matrix, correlated-Wishart matrix, and their weighted sum as given in Eq.~\eqref{SOW2}. The parameter values are $n=6,n_A=n_B=9$, $a=b=1$, and $(\sigma_1,...,\sigma_6)=(4,20/3,5/2,11/9,4/3,7/8)$. }
  \label{Fig8}
\end{figure}
% FIGURE 8

% SECTION 6
\section{Weighted sum of two Wisharts}
\label{Sec6}

%SUBSECTION 6a
\subsection{Matrix model and probability density}
We finally consider the matrix model
\begin{equation}
\label{SOW2}
H=a A+b B,
\end{equation}
where $A$ and $B$ are $n$-dimensional positive-definite-Hermitian matrices, respectively, from the distributions
\begin{equation}
\mathcal{P}(A)\propto e^{-\tr A}|A|^{n_A-n},~~~\mathcal{P}(B)\propto e^{-\tr \Sigma^{-1} B}|B|^{n_B-n},
\end{equation}
with $n_A,n_B\ge n$. The parameters $a,b$ are again non-negative scalars.
We have taken the covariance matrix equal to identity matrix for $A$, while for $B$ we have assumed an arbitrary (positive definite) covariance matrix. The latter constitutes the correlated variant of the Wishart ensemble. The above matrix model has been considered in~\cite{Kumar2014} and exact results have been obtained for the matrix probability density, the joint probability density of eigenvalues, as well as the first order marginal density. 

We would like to remark that if one considers covariance matrices proportional to identity matrix only, then the problem can be solved for the weighted sum of arbitrary number of Wishart matrices. Such a scenario has been considered in~\cite{KGGC2015} and the results used for the analysis of multiuser communication employing multiantenna elements such as multiple-input multiple-output (MIMO) multiple access channel (MAC). 

For the matrix model of Eq.~\eqref{SOW2}, with parameter $m=n_A+n_B-n$, the probability density function satisfied by matrix $H$ reads~\cite{Kumar2014}
\begin{align}
\nonumber
&\label{PH}\mathcal{P}_H(H)\propto|H|^m\, e^{-\tr(a^{-1} H)} ~~~~~\\
&\times\,_1\boldsymbol{F}_1(n_B;n_A+n_B;(a^{-1}\1-b^{-1}\Sigma^{-1})H),
\end{align}
where $ \,_1\boldsymbol{F}_1$ is confluent hypergeometric function of the first kind (Kummer's function) with matrix argument:
\begin{align}
\nonumber
& \,_1\boldsymbol{F}_1(\alpha,\gamma;X)=\frac{1}{B_n(\alpha,\gamma-\alpha)}\\
 &\times\int_0^{\1} \!\!d[Y]e^{\tr XY}|\1-Y|^{\alpha-n}|Y|^{\gamma-\alpha-n}.
\end{align}
Here $B_n(\alpha,\gamma)$ is the multivariate beta function:
\begin{align}
B_n(\alpha,\gamma)=\int_0^{\1} \!\!d[Y]|\1-Y|^{\alpha-n}|Y|^{\gamma-n}.
\end{align}
Similar to the beta function with scalar arguments, it is related to multivariate gamma function in Eq.~\eqref{multigamma} as
\begin{equation}
\label{multibeta}
B_n(\alpha,\gamma)=\frac{\Gamma_n(\alpha)\,\Gamma_n(\gamma)}{\Gamma_n(\alpha+\gamma)}.
\end{equation}

%SUBSECTION 6b
\subsection{Eigenvalue statistics}
The joint probability density of eigenvalues ($0<\lambda_1,...,\lambda_n<\infty$) for Eq.~\eqref{SOW2} is given by Eq.~\eqref{biortho} with
\begin{equation}
w(\lambda)= \lambda^{m} e^{- \lambda/a},
\end{equation}
\begin{equation}
f_j(\lambda_k)=\,_1F_1\Big(n_B-n+1;\,m+1;\, \Big(\frac{1}{a}-\frac{1}{b\sigma_{j}}\Big) \lambda_k\Big),
\end{equation}
where $\sigma_j$ are the eigenvalues of $\Sigma$~\cite{Kumar2014}.
Also, $h_{j,k}$ can be obtained using the result
\begin{equation}
\label{Int}
\int_0^\infty\!\!d\lambda\,  \lambda^m e^{-s \lambda} \,_1F_1(a; b; c \lambda)=\frac{\Gamma(m+1)}{s^{m+1}}\!\,_2F_1\Big(a;m+1;b;\frac{c}{s}\Big),
\end{equation}
valid for convergent scenarios, as
\begin{align}
\nonumber
&h_{j,k}=a^{m+k}\,\Gamma(m+k)\\
&\times\,_2F_1\left(n_B-n+1;m-n+k;m+1;1-\frac{a}{b\sigma_j}\right).
\end{align}
Consequently, we obtain an explicit result for the $r$-point correlation function.

Figure~\ref{Fig7a} shows the two-point correlation function of eigenvalues for matrix model given in Eq.~\eqref{SOW2}, while Fig.~\ref{Fig7b} depicts the marginal density. In Fig.~\ref{Fig8} we compare the densities of Wishart, correlated-Wishart and their weighted sum. For density of the correlated-Wishart we have used the following result, which also follows with the aid of Eq.~\eqref{marginal}:
\begin{align}
\nonumber
&p_{\mathrm{CW}}(\lambda)=-\frac{\lambda^{s}}{n\Delta(\{\sigma\})\prod_{j=1}^n\sigma_j^{s+1}\Gamma(j+s)} \\
&\times\begin{vmatrix} 0 &  [\lambda^{k-1}]_{k=1,...,n}  \\  [e^{-\sigma_j^{-1}\lambda}]_{j=1,...,n} & [\Gamma(k+s)\sigma_j^{k+s}]_{\substack{j=1,...,n\\k=1,...,n}} \end{vmatrix},
\end{align}
where $s=n_B-n$. We note that while matrix models~\eqref{ratio} and~\eqref{SOW2} can lead to similar densities in particular scenarios (e.g. when $b\rightarrow 0$), in general they exhibit different behavior. For instance, the present model can not lead to the multivariate-beta-distribution kind of density which can be achieved using Eq.~\eqref{ratio} for $a=b\rightarrow\infty$.

% SECTION 7
\section{Conclusion and Outlook}
\label{Sec7}

We considered four important matrix models which lead to biorthogonal structure in their joint eigenvalue densities. These matrix ensembles play important roles in several areas, which range from multiple antenna communication theory to supergravity theory. We evaluated the matrix distribution, as well as the joint eigenvalue density for these ensembles. With the information of joint density, we also presented determinantal expression for eigenvalue correlation function of arbitrary order. This representation follows from a generalization of Andr\'{e}ief's integration formula. Since knowledge of the correlation function gives access to the prediction of statistical behavior of observables of interest in a given problem, we believe that the exact results derived here will find interesting applications in several fields.

As continuation of the present work, an immediate direction to pursue could be the investigation of the behavior of extreme eigenvalues of the composite matrix models, and its comparison with that of the extreme eigenvalues of the constituent matrices. This will give a better insight into the mechanism by which the redistribution of eigenvalues takes place. Since all the matrix models considered here possess biorthogonal structure in their joint eigenvalue density expressions, exact results are possible for the gap probabilities and densities of extreme eigenvalues ~\cite{Kumar2015}.

While we considered here ensembles comprising complex matrices, the cases of real and quaternion matrices are also important and can be explored. However, solving these ensembles poses serious challenges because of unavailability of group integral results similar to those in the case of unitary group. Another interesting direction can be the analysis of the spectra of the composite matrices in large dimension limit, and to look for universalities.

%APPENDICES
\appendix

% Appendix 1
\section{Correlation function}
\label{AppA}
We will use mathematical induction to prove Eq.~\eqref{corrfunc}. Equations~\eqref{corrdef} and~\eqref{corrfunc} are defined for $r=1,2,...,n$~\footnote{ One may define $R_0(\_)$ being equal to 1.}. From the definition of correlation function, Eq.~\eqref{corrdef}, it is clear that
\begin{equation}
\label{recur}
R_{r-1}(\lambda_1,...,\lambda_{r-1})=\frac{1}{n-r+1}\int d\lambda_r\,R_r(\lambda_1,...,\lambda_r).
\end{equation} 
For $r=n$ Eq.~\eqref{corrfunc} clearly holds, since in this case the determinant in Eq.~\eqref{corrfunc}  factorizes into the product of two determinants and produces $n!\,P(\{\lambda\})$. Let us assume it is correct for $r=s$. We will prove that given this,  Eq.~\eqref{corrfunc} holds for $r=s-1$ as well.
\begin{widetext}
Using Eq.~\eqref{recur} we obtain
\begin{equation}
\label{induct}
R_{s-1}(\lambda_1,...,\lambda_{s-1})=\frac{(-1)^s n!\,C}{n-s+1} \prod_{l=1}^{s-1} w(\lambda_l)\int d\lambda_s w(\lambda_s)\begin{vmatrix}    [0]_{\substack{j=1,...,s\\k=1,...,s}} &  [\lambda_j^{k-1}]_{\substack{j=1,...,s\\k=1,...,n}}  \\  [f_j(\lambda_k)]_{\substack{j=1,...,n\\k=1,...,s}}  & [h_{j,k}]_{\substack{j=1,...,n\\k=1,...,n}} \end{vmatrix}.
\end{equation} 
We expand the determinant using the $s$'th row:
\begin{equation}
\begin{vmatrix}    [0]_{\substack{j=1,...,s\\k=1,...,s}} &  [\lambda_j^{k-1}]_{\substack{j=1,...,s\\k=1,...,n}}  \\  [f_j(\lambda_k)]_{\substack{j=1,...,n\\k=1,...,s}}  & [h_{j,k}]_{\substack{j=1,...,n\\k=1,...,n}} \end{vmatrix}
=\sum_{\mu=1}^n(-1)^{2s+\mu}\lambda_s^{\mu-1}\begin{vmatrix}   [0]_{\substack{j=1,...,s-1\\k=1,...,s}}  &  [\lambda_j^{k-1}]_{\substack{j=1,...,s-1\\k=1,...,n\\(k\neq \mu)}} \\  [f_j(\lambda_k)]_{\substack{j=1,...,n\\k=1,...,s}}   & [h_{j,k}]_{\substack{j=1,...,n\\k=1,...,n\\(k\neq \mu)}} \end{vmatrix}.
\end{equation}
We now insert the $w(\lambda_s)\lambda_s^{\mu-1}$ in the $s$'th column, and perform the $\lambda_s$ integral. Using the definition of $h_{j,k}$ given in Eq.~\eqref{hjk} , we obtain 
\begin{equation}
\sum_{\mu=1}^n(-1)^{\mu}\begin{vmatrix}   [0]_{\substack{j=1,...,s-1\\k=1,...,s-1}} & [0]_{j=1,...,s-1}  &  [\lambda_j^{k-1}]_{\substack{j=1,...,s-1\\k=1,...,n\\(k\neq \mu)}}  \\  [f_j(\lambda_k)]_{\substack{j=1,...,n\\k=1,...,s-1}} & [h_{j,\mu}]_{j=1,...,n}   & [h_{j,k}]_{\substack{j=1,...,n\\k=1,...,n\\(k\neq \mu)}} \end{vmatrix}.
\end{equation}
Performing separate row interchanges in the determinants appearing in the sum, we arrive at
\begin{equation}
\label{detsum}
\sum_{\mu=1}^n(-1)^{2\mu-1}\begin{vmatrix}   [0]_{\substack{j=1,...,s-1\\k=1,...,s-1}}  & [(1-\delta_{\mu,k})\lambda_j^{k-1}]_{\substack{j=1,...,s-1\\k=1,...,n}}  \\
[f_j(\lambda_k)]_{\substack{j=1,...,n\\k=1,...,s-1}} & [h_{j,k}]_{\substack{j=1,...,n\\k=1,...,n}} \end{vmatrix},
\end{equation}
where $\delta_{\mu,\nu}$ is the Kronecker-delta function.
\end{widetext}
Using multilinearity property in first $s-1$ rows in determinant appearing in each of the terms in the above summation, we find that it gives rise to
\begin{align}
(-1)^{-1}(n-s+1)\begin{vmatrix}   [0]_{\substack{j=1,...,s-1\\k=1,...,s-1}}  & [\lambda_j^{k-1}]_{\substack{j=1,...,s-1\\k=1,...,n}}  \\
[f_j(\lambda_k)]_{\substack{j=1,...,n\\k=1,...,s-1}} & [h_{j,k}]_{\substack{j=1,...,n\\k=1,...,n}} \end{vmatrix}.
\end{align}
Plugging this back in Eq.~\eqref{induct}, we obtain an expression for $R_{s-1}(\lambda_1,...,\lambda_{s-1})$ which is consistent with Eq.~\eqref{corrfunc}, and hence the desired result follows.

% Appendix 2
\section{Matrix Integral}
\label{AppB}

Consider $n$-dimensional Hermitian matrices $X$ and $Y$. We are interested in evaluating integral of the form
\begin{equation}
\label{Laplace}
\Phi(X)=\int d[Y]\,e^{-s \tr X Y} F(Y),
\end{equation}
where $s$ is a scalar and $F(Y)$ is a unitarily invariant expression involving $Y$, such that the above integral is convergent. We note that Eq.~\eqref{Laplace} is a matrix generalization of Laplace transform. If $\boldsymbol{x}$ and $\boldsymbol{y}$ are the diagonal matrices consisting of eigenvalues of $X$ and $Y$, then
\begin{align}
\nonumber
\Phi(X)=\int_0^\infty dy_1\cdots \int_0^\infty dy_n \Delta^2_n(\{y\})F(\boldsymbol{y})\\
\times\int_{\mathcal{U}_n} d\mu(\mathcal{U})e^{-s \tr(\boldsymbol{x}\,\mathcal{U}^\dag\, \boldsymbol{y}\,\mathcal{U})},
\end{align}
where $d\mu(\mathcal{U})$ represents the Haar measure over the group  $\mathcal{U}_n$ of $n$-dimensional unitary matrices.
The unitary group integral can be performed using the celebrated Harish-Chandra--Itzykson-Zuber formula~\cite{HC1956,IZ1980} and leads to
\begin{align}
\nonumber
\Phi(X)\propto\frac{1}{\Delta_n(\{x\})}\int_0^\infty dy_1\cdots \int_0^\infty dy_n\Delta_n(\{y\})\\
\times F(\boldsymbol{y})\,\left|e^{-s x_j y_k}\right|_{j,k=1,...,n} .
\end{align}
Now if $F(\boldsymbol{y})$ is expressible in terms of certain weight functions $u(y_j)$ as $F(\boldsymbol{y})=\prod_{j=1}^n u(y_j)$, then integral over $\boldsymbol{y}$ can be performed and results in
\begin{equation}
\Phi(X)\propto\frac{1}{\Delta_n(\{x\})} \left|f_j(x_k)\right|_{j,k=1,...,n} ,
\end{equation}
where
\begin{equation}
f_j(x_k)=\int_0^\infty dy\, u(y)\, y^{j-1}\,e^{-s\, x_ky}.
\end{equation}
Note that we may consider $j\rightarrow n-j+1$ (or/and $k\rightarrow n-k+1$) for $f_j(x_k)$ within the determinant in~\eqref{biortho} and then accordingly modify the rest of the results in Sec.~\ref{Sec2} which depend on $f_j(x_k)$.

% REFERENCES
%\section*{References}

\end{document}